%% file: paper_after_review_2.tex
\documentclass[10pt,twocolumn,final]{IEEEtran}

\usepackage{amsmath, mathrsfs}
\usepackage{amsfonts}
\usepackage{amssymb}
\usepackage{graphicx}
\usepackage{color}
\usepackage{multirow}
\usepackage{graphicx}

\usepackage{algorithm}
\usepackage{algpseudocode}
\usepackage{pifont}
\usepackage{varwidth}

\usepackage[
	a4paper,
        left=1.7cm,
        right=1.7cm,
        top=1.95cm,
        bottom=1.95cm]
{geometry}

\newcommand{\myhash}{%
  {\settoheight{\dimen0}{C}\kern-.05em\, \resizebox{!}{\dimen0}{\raisebox{\depth}{\#}}}}

\usepackage{caption}
\usepackage{subcaption}

\usepackage[numbers,sort&compress]{natbib}

\usepackage{multicol}


\input{symbols.txt}

\usepackage{graphicx}
\usepackage{tikz}
\usepackage{pgfplots}
\pgfplotsset{compat=newest}
\usetikzlibrary{patterns}
\usetikzlibrary{positioning}
\usetikzlibrary{datavisualization}
\usetikzlibrary{datavisualization.formats.functions}
\usetikzlibrary{backgrounds}
\usetikzlibrary{shapes,snakes}

\usepackage{float}
\usepackage{afterpage}
\usepackage{balance}

\renewcommand\thesection{\arabic{section}}
\renewcommand\thesubsection{\thesection.\arabic{subsection}}
\renewcommand\thesubsubsection{\thesubsection.\arabic{subsubsection}}

\renewcommand\thesectiondis{\arabic{section}}
\renewcommand\thesubsectiondis{\thesectiondis.\arabic{subsection}}
\renewcommand\thesubsubsectiondis{\thesubsectiondis.\arabic{subsubsection}}

\def\herm{{\sfH}}

\def\snr{{\mathsf{snr}}}

\def\cg{{\clC\clN}} 
\newcommand{\up}[1]{_{#1}}

\input{macros.txt}

\begin{document}

\title{Enhancing the Estimation of mm-Wave Large Array Channels by Exploiting
Spatio-Temporal Correlation and Sparse Scattering}
\author{Saeid Haghighatshoar,  \IEEEmembership{Member, IEEE,} Giuseppe Caire,
\IEEEmembership{Fellow, IEEE}%
\thanks{The authors are with the Communications and Information Theory Group, Technische Universit\"{a}t Berlin (\{saeid.haghighatshoar, caire\}@tu-berlin.de).}
\vspace{-2mm}}

\maketitle

\begin{abstract}
In order to cope with the large path-loss exponent of mm-Wave channels, a high beamforming gain is needed. 
This can be achieved with small hardware complexity and high hardware power efficiency
by Hybrid Digital-Analog (HDA) beamforming, 
where a very large number $M \gg 1$ of antenna array elements requires 
only a relatively small $m \ll M$ number of A/D converters and modulators/demodulators. 
As such, the estimation of mm-Wave MIMO channels must deal with two specific problems: 
1) high Doppler, due to the large carrier frequency; 2) impossibility of observing directly the $M$-dimensional channel vector at the antenna array elements, 
due to the mentioned HDA implementation. 
In this paper, we consider a novel scheme inspired by recent results on gridless multiple measurement vectors problem in 
compressed sensing, that is able to exploit the inherent mm-Wave channel sparsity in the angular domain in order to 
cope with both the above problems simultaneously.  Our scheme uses past pilot-symbol observations in a window of length $T$ 
in order to estimate a low-dimensional subspace that approximately contains the channel vector at the current time. 
This subspace information can be used directly, in order to separate users in the spatial domain, or indirectly,  
in order to improve the estimate of the user channel vector from  the current pilot-symbol observation.
\end{abstract}

\section{Introduction}
Millimeter wave (mm-Wave) communication is a promising technology for the next generation of 
WLANs and outdoor cellular systems \cite{pi2011introduction,rappaport2013millimeter}. 
In order to cope with the large path-loss exponent of mm-Wave channels, a high beamforming gain is needed. 
While large antenna arrays can be implemented with a small form factor due to the small wavelength, 
it is clear that conventional all-digital baseband processing as proposed for large MIMO systems at lower frequencies
\cite{Marzetta-TWC10,shepard2012argos,larsson2014massive}  is not a suitable solution here. 
In fact, because of the large signal bandwidth available at mm-Waves, the demodulation and quantization of the signal at each antenna array 
element would require an enormous A/D front-end bit-rate, with corresponding unacceptable hardware power consumption. 
For this reason, a promising approach for mm-Wave communication is the 
Hybrid Digital-Analog (HDA) beamforming, where the beamforming function is achieved in two stages. 
The first stage uses as analog reconfigurable beamforming network operating in the RF domain, and achieves beamforming gain and
some coarser multiuser interference rejection while reducing the signal dimension from $M \gg 1$ (number of antenna array elements) to 
some $m \ll M$ (number of RF chains and A/D converters). The second stage, processes the $m$-dimensional baseband signal 
in the digital domain in order to achieve further multiuser MIMO spatial multiplexing gain \cite{adhikary2013joint,adhikary2014joint}.

For multiuser spatial multiplexing, the base station needs to estimate the $M$-dimensional channel vectors of all the users. Channel estimation for mm-Wave MIMO channels must deal with two specific problems: 
1) potentially rapid variations of the small-scale fading coefficients, due to the large carrier frequency; 
2) impossibility of observing directly the $M$-dimensional channel vectors of the users at the antenna array elements, 
due to the mentioned HDA implementation. 
Fortunately, mm-Wave channels have a special feature that helps to cope with both the above problems simultaneously, namely,  
the resulting channel vectors are typically very sparse in the angular domain, since only the Line-of-Sight path and/or a few dominant multipath components
convey  significant  power\footnote{This is especially true in the case of a tower-mounted base-station and/or in the case of mm-wave channels, as experimentally confirmed by 
channel measurements (see \cite{adhikary2014joint} and references therein).}.  

In this paper, we consider a novel scheme inspired by recent results on gridless multiple measurement vectors problem in 
compressed sensing, that exploits the inherent mm-Wave channel sparsity in the angular domain in order to 
cope with both the above problems. In this scheme, we exploit the past pilot-symbol observations in a window of length $T$ 
in order to estimate a low-dimensional subspace that approximately contains the channel vector at the current time slot. 
This subspace information can be used directly, to separate users in the spatial domain, or indirectly,  
 to improve the estimate of the user channel vector in the current time slot. 
 Simulations show very encouraging preliminary results, and in particular confirm that the 
channel subspace information obtained over a window of past measurements provides significant improvements 
with respect to the conventional ``one-shot'' techniques, that estimate the channel vectors by using only the current pilot 
observation. 

\noindent {\bf Notations:}
We denote vectors by boldface small letters (e.g., $\xv$), matrices by boldface capital letters (e.g., $\Xm$), scalar constant by 
non-boldface letters (e.g., $x$ or $X$), and sets by calligraphic letters (e.g., $\Xc$). The $i$-th element of a vector $\xv$
and the $(i,j)$-th element of a matrix $\Xm$ will be denoted by $[\xv]_i$ and $[\Xm]_{i,j}$ respectively. 
We denote the Hermitian and the transpose of a matrix  $\bfX$ by $\bfX^\herm$ and $\Xm^\transp$, respectively. 
The same notation is used for vectors and scalars. 
We use $\bT_+$ for the space of Hermitian semi-definite Toeplitz matrices. For an $\bfx \in \bC^M$, we denote by $\bT(\bfx)$ a Hermitian Toeplitz matrix whose first column is $\bfx$.
We always use $\Id$ for the identity matrix, where the dimensions may be explicitly indicated for the sake of clarity (e.g., $\Id_p$ denotes
the $p \times p$ identity matrix).   
For an integer $k \in \ZZ$, we use the shorthand notation $[k]$ 
for the set of non-negative integers $\{0,1,2, \dots, k-1\}$, where the set is empty if $k < 0$. 

\section{Model and Problem Statement}

\subsection{Channel Model}\label{sec:ch_model}

Motivated by mm-Wave channel measurements and models \cite{rappaport2013millimeter}, 
we consider a simple propagation model for the wireless scattering channel in which the transmission between a single-antenna user and 
the $M$-antenna base-station array occurs through $p \ll M$ multipath components (see Fig.~\ref{fig:sc_channel}). 
The base-station is equipped with a {\em Uniform Linear Array} (ULA), with spacing $d=\frac{\lambda}{2 \sin(\theta_{\max})}$ between its elements, 
with $\lambda$ being the wave-length, and scans the angular range $[-\theta_{\max}, \theta_{\max}]$ for some $\theta_{\max} \in (0,\pi/2)$. 
We denote by $\bfa(\theta)\in \bC^M$ the array response for the AoA $\theta \in [-\theta_{\max}, \theta_{\max}]$, 
whose $k$-th component, is given by  $[\bfa(\theta)]_k=e^{jk\frac{2\pi d}{\lambda} \sin(\theta)}= e^{jk\pi \frac{\sin(\theta)}{\sin(\theta_{\max})}}$.
\input{channel_fig.txt}
We consider a discrete-time model, where the channel vector of a user at time $t$ 
is given by 
\begin{align}\label{eq:h_t}
\bfh[t]=\sum_{\ell=1}^{p} w_\ell[t] \bfa(\theta_\ell),
\end{align}
where $\theta_\ell$ denotes the angle-of-arrival (AoA) of the  $\ell$-th multipath component and 
where $w_\ell[t]$ is the corresponding small-scale fading coefficient, assumed $\sim \cg(0, \sigma_\ell^2)$. 
According to the well-known Wide-Sense Stationary Uncorrelated Scattering (WSSUS) model, the coefficients
$w_\ell[t]$ are WSS processes with respect to $t$ and mutually uncorrelated with respect to $\ell$. 
The general wisdom of multiuser MIMO considers ``one-shot'' or ``instantaneous'' estimation \cite{Marzetta-TWC10}. 
This consists of partitioning the slot into a training phase and
a data transmission phase. The channel vectors are estimated during the training phase, and these estimates are used in the data transmission phase. 
In compliance with most of the recent ``massive MIMO'' literature \cite{larsson2014massive},  
we assume Time-Division Duplexing (TDD) and channel reciprocity \cite{shepard2012argos}, such that 
the channel vectors of the users are estimated during a training phase, in which orthogonal (uplink) pilot symbols are transmitted by the users to the base-station. The resulting estimates are used in data transmission phase
to receive data streams transmitted simultaneously by the users 
to the base-station (uplink), or to transmit multiple data streams from the
 base-station to the users (downlink).
In both cases, the data streams are separated  in the spatial domain by linear beamforming (spatial multiplexing). 

As anticipated in the introduction, in mm-Wave channels the ``instantaneous'' channel estimation may suffer from the fact that the mm-Wave channels change rapidly in time. Therefore, the ability of the beamformer to eliminate the multiuser interference in the spatial domain may be impaired by the ``channel aging'' phenomenon, i.e.,  by the time the channel estimate is used, the channel has already significantly changed. 
In addition, due to the discussed HDA implementation of the base-station front-end, 
the whole $M$-dimensional received signal in correspondence of the uplink pilot symbols cannot be fully observed. 
Rather, only an $m$-dimensional projection (or ``sketch'') through the analog beamforming network (consisting of $m$ separate RF chains) is available. 

While the channel vectors may change rapidly in time (up to the limit of having i.i.d. channels across different time slots), 
the WSS assumption implies that the scattering geometry, expressed by the AoA's $\{\theta_\ell\}_{\ell=1}^p$ and the multipath component strengths 
$\{\sigma_\ell^2\}_{\ell=1}^p$, remains invariant for a very large number of slots. This is justified by the fact that 
the ``small-scale fading'' channel gains $w_\ell[t]$ go through a full phase cycle when the distance between transmitter and receiver varies by 
one wavelength (e.g., 1cm at 30 GHz), whereas  AoAs and path strengths change only when the ``large-scale'' geometry of the propagation between the transmitter  and the receiver significantly changes.\footnote{Strictly speaking, according to the widely accepted Wide-Sense Stationary Uncorrelated Scattering (WSSUS) model,  the second-order statistics of the channel vector process are time-invariant, implying that AoAs and signal strengths are strictly constant in time. 
As a mater of fact, the WSSUS model is a local approximation, with coherence time much larger than the small-scale fading coherence time.}

In this paper, we consider on a TDD scenario, where the users send their pilot signals in non-overlapping time intervals, thus, without loss of generality, we can focus on the channel estimation problem of an individual user. We assume that the uplink pilot symbols for the user are sent 
periodically with a period $\tau$, and is accumulated in an observation window of $T$ slots, thus, in total there are $\nu=\frac{T}{\tau}$ training 
samples (see Fig.~\ref{fig:pilot}). 
\input{training_fig.txt}
The received signal at the $i$-th training period, $i \in [\nu]$, is given by $\bfy\up{i}=\bfh\up{i} s_i + \bfn\up{i}$, where $\bfh\up{i}=\bfh[i\tau]$ denotes the random channel vector of the user (at time $t=i\tau$), where $s_i$ denotes the $i$-th training symbol, and where $\bfn\up{i} \sim \cg(0, \sigma^2\bfI)$ is the additive white Gaussian noise of the array. For simplicity, we will assume that $s_i=1$ for the rest of the paper.
We define the training \textit{signal-to-noise-ratio} (SNR) by $\snr=\sum_{\ell=1}^p \sigma_\ell^2 / \sigma^2$. 
Once an estimate of the channel vector $\bfh\up{i}$ is available,  it is used in the data transmission phase of the current slot $\clT_i = [i\tau, (i+1)\tau - 1]$ 
to calculate the beamformer for the base-station receiver (uplink) and/or the base-station transmitter (downlink).

\subsection{One-Shot Sparse Channel Estimation}\label{sec:ch_estim}

Since by assumption we have $p \ll M$,  the channel vector $\bfh\up{i}$, $i \in [\nu]$, has a sparse representation in the continuous dictionary $\clA=\{\bfa(\theta): \theta \in [-\theta_{\max}, \theta_{\max}]\}$ consisting of the array responses for different AoAs $\theta$,  with the sparsity being $\frac{p}{M} \ll 1$. 
Classical compressed sensing (CS) methods \cite{donoho2006compressed,candes2006near} can be used to estimate $\bfh\up{i}$ via a few, say $m\ll M$, linear projections 
of the received signal $\bfy\up{i}$ rather than the whole components thereof. This feature is well-suited for the HDA front-end implementation that supports a number of RF chains and A/D converters much smaller than
the number of array elements.  Let us denote the $m\times M$ measurement projection matrix by $\bfB$, 
where we assume that the rows of $\bfB$ are orthonormal\footnote{Since $\Bm$ is the projection matrix corresponding to the RF 
beamforming receiver, it can be designed to satisfy row orthonormality.}. 
Also let $\bfx\up{i}=\bfB \bfy\up{i}= \bfB(\bfh\up{i} +\bfn\up{i})$, $i \in [\nu]$, be the resulting $m$-dimensional projections. 
To recover the sparse signal $\bfh\up{i}$, we use the atomic-norm denoising algorithm \cite{chandrasekaran2012convex}
\begin{align}\label{atomic_denoising}
\widehat{\bfh}\up{i}=\argmin \|\bfh\|_\clA \text{ s.t. } \|\bfx\up{i} - \bfB \bfh\|^2 \leq \epsilon,
\end{align}
where $\epsilon\approx m\sigma^2$ is an estimate of the noise power, and where $\|\bfh\|_\clA$ denotes the atomic norm of $\bfh$ with respect to the continuous dictionary of the array vectors $\clA$, defined by
\begin{align}
\|\bfh\|_\clA=\inf \Big \{\sum_\ell &c_\ell: c_\ell \geq 0, \text{and }\nonumber\\
 &\exists\,(\theta_\ell, \phi_\ell) \text{ s.t. } \bfh=\sum_\ell c_\ell e^{j \phi_\ell} \bfa(\theta_\ell)\Big \}.
\end{align}
In general, finding a closed-form formula or even efficiently computing the atomic norm of a vector in a given dictionary is a challenging task, and different methods have been proposed for its approximation \cite{chandrasekaran2012convex}. However, for the dictionary $\clA$, it has been shown that the atomic norm can be efficiently computed 
via \textit{semi-definite programming} (SDP) \cite{bhaskar2013atomic}. Thus, the atomic-norm denoising \eqref{atomic_denoising} for estimating the sparse 
channel vector $\bfh\up{i}$ can be written as the following SDP:
\begin{align}\label{sparse_alg}
\widehat{\bfh}\up{i}=&\argmin_{\bfh\in \bC^M, \bfv\in \bC^M, \gamma\in \bR_+} \trace[\bT(\bfv)]+ \gamma \text{ s.t. }\nonumber\\
&\left [ \begin{array}{cc} \bT(\bfv) & \bfh\\ \bfh^\herm & \gamma \end{array} \right ] \succeq {\bf 0}, \|\bfx\up{i} - \bfB \bfh\|^2 \leq \epsilon,
\end{align}
where $\bT(\bfv)$ denotes an $M \times M$ Hermitian Toeplitz matrix whose first column is $\bfv$, and where $\epsilon=m\sigma^2$ is an estimate of the noise power. 

In this paper, we will use optimization \eqref{sparse_alg} as the \textit{one-shot} sparse channel estimation algorithm since it uses only the 
observation $\bfx\up{i}$ on the current slot $i$ and does  not exploit the previous training samples in a window of duration $\nu$ consisting of  $\big\{\bfx\up{j} : j \in \{i-\nu, i-\nu+1, \ldots, i - 1\}\big \}$.

\subsection{Time Variation of the Channel Vectors}\label{sec:ch_var}

For the sake of simplicity, we assume that the multipath component coefficients
evolve according to first order Markov processes given by
\begin{align}\label{fo_markov}
w_\ell[t]=\alpha_\ell\,w_\ell[t-1] + \sigma_\ell \sqrt{1-\alpha_{\ell}^2}\ i_\ell[t],
\end{align}
where $i_\ell[t]$ is the innovation process for $w_\ell[t]$, which is a Gaussian process with a covariance $\bE[i_\ell[t] i_{\ell'}[t']]=\delta_{\ell, \ell'} \delta_{t,t'}$, and where $\alpha_\ell$  is the coefficient of first order auto-regression filter, which should be inside the unit circle to have a stable filter, i.e., $|\alpha_\ell|<1$. 
To obtain a stationary process, we assume that $w_\ell[0]\sim \cg(0, \sigma_\ell^2)$ is initialized with the first realization of the channel 
gain for the $\ell$-th scatterer. In this case, $w_\ell[t]$ generated by \eqref{fo_markov} is a stationary Gaussian process for all $t\geq 0$, whose auto-correlation function is given by
\begin{align}
r_\ell[\Delta]=\bE\big[w_\ell[t+\Delta] w_\ell[t]^*\big]=\sigma_{\ell}^2\,\alpha_\ell^{|\Delta|}.
\end{align}
For simplicity, we assume that $\alpha_\ell=\alpha$ is the same for all $\ell$, and $\alpha \in [0,1)$ is real-valued and positive. Since $\bfh\up{i}$ is obtained by sampling $\bfh[t]$ every $\tau$ seconds, the matrix-valued 
auto-correlation function  of $\bfh\up{i}$ is given by
\begin{align}\label{auto_corr_h}
\bE[\bfh\up{i} \bfh\up{i'}^\herm]
&= \beta^{|i-i'|} \sum_{\ell=1}^p \sigma_\ell^2 \bfa(\theta_\ell) \bfa^\herm(\theta_\ell)
\end{align} 
where $\beta=\alpha^{\tau}$.
Without loss of generality, we shall consider a measurement window $[\nu] = \{0, \ldots, \nu-1\}$ of $\nu$ slots, and look at the transmitter/receiver operations in slot 
$\clT_\nu=[\nu \tau, (\nu+1)\tau-1]$. Therefore, the measurement window is referred to as a block of ``past observations'', while the measurement at slot $\nu$ is the ``current observation''. We define the \textit{coherence time} (or the settling time) of the channel \cite{tse2005fundamentals} by $\tau_c=\frac{1}{\log (1/\alpha)}$.
We consider three idealized cases of interest: 
\begin{enumerate}
\item When $T\ll \tau_c$, the channel process is almost constant over a time significantly larger than $T$. It follows that the channel on the current slot is approximately identical to
the channel over the whole past observation window. In this case, predicting the channel on the current slot from the past window is expected to be very effective. 
\item When $\tau \ll  \tau_c \approx T$, the channel varies significantly over the past observation window, but remains approximately constant over each slot.
Hence, one-shot estimation over the current slot yields an accurate estimate in high SNR. However, since channel estimation is performed {\em before} beamforming (in fact, it is used
to calculate the beamformer) in mm-Wave communication it is reasonable to expect that estimation occurs in low SNR (without array beamforming gain). 
Hence, we are interested in using the past observation window to {\em improve} the one-shot estimation of the current channel. 
\item When $\tau_c \approx \tau$, the channel process varies significantly over a slot (i.e., it is nearly i.i.d. over different slots). 
In this case, one-shot estimation is ineffective due to channel aging over the current slot, 
especially in the donwlink case. Nevertheless, we can learn the channel dominant subspace, i.e., the linear span of the atoms that best represent the channel over the past observation window, and still be able to  separate the users in the signal space based only on subspace information. This is effective when such channel subspaces are low-dimensional, as is the case for mm-Wave  channels \cite{adhikary2014joint}. 
\end{enumerate}

\subsection{Exploiting Past Measurements}\label{exp_past_meas}

In order to illustrate the fact that both sparsity in the AoA domain and time correlation can be used
to improve channel estimation, we consider two extremes of cases 1) and 3) said above.
In the first case, the channel is exactly constant over an interval much larger than $T$, i.e., 
$\bfh[i\tau]=\bfh\up{i}=\bfh\up{0}$ for $i \in [\nu]$, where $h[t]$ is given by \eqref{eq:h_t}. Hence, by simply averaging the training observations $\bfx\up{i}=\bfB \bfy_i$ for $i \in [\nu]$, we obtain  
\begin{align}  \label{time-avg}
\overline{\bfx}=\frac{1}{\nu} \sum_{i\in [\nu]} \bfx_i = \bfB \left (\bfh\up{0} + \frac{1}{\nu} \sum_{i \in [\nu]} \bfn_i \right ).
\end{align}
Applying the one-shot sparse estimator (\ref{sparse_alg}) to (\ref{time-avg}), we obtain an estimate of $\hv_\nu \approx \hv_0$ with an improvement in the observation SNR by a factor of $\nu$. Furthermore, because of the strong correlation in time, the system does not even need to exploit the observation on the current slot (this would only improve the SNR by a marginal factor of $(1 + 1/\nu)$). This means that, for highly time-correlated channel dynamics, 
channel prediction can be effectively exploited. 

Now consider the opposite extreme case,  where the channel gains are i.i.d. over the sequence of slots. 
Let us consider the sample covariance estimator $\widehat{\bfC}_x=\frac{1}{\nu} \sum_{i \in [\nu]}\bfx\up{i} \bfx\up{i}^\herm$. 
By the consistency of the sample covariance, for sufficiently large $\nu$, we have
\begin{align}
\widehat{\bfC}_x \approx \bfB \bfC_h \bfB^\herm + \sigma^2 \bfB \bfB^\herm=\bfB \bfC_h \bfB^\herm + \sigma^2 \bfI_m,
\end{align}
where we have assumed that the rows of $\bfB$ are orthonormal.
In our previous work \cite{haghighatshoar2015channel}, we showed that it is possible to exploit the angular sparsity and the underlying Toeplitz structure 
of $\bfC_h$ (for the ULA), such that the $p$-dimensional signal subspace that contains $\bfh\up{i}$ with probability $1$, namely, Span$\{ \av(\theta_\ell) : \ell = 1,\ldots, p\}$, 
 be efficiently estimated when the projection matrix $\Bm$ has 
only $m\approx 2\sqrt{M}$ rows. As a matter of fact, it is sufficient to let $\Bm$  have a single non-zero element equal to 1 in each row, such that 
$\Bm$ induces a subsampling of the array elements (antenna selection) in coprime locations. 
In particular, $\nu$ of the order $\sim 50-100$ samples seems to be sufficient to precisely estimate this subspace for 
moderate SNR values around $\snr \sim 0-10$ dB. 

Let $\bfU$ be the $M \times p$ tall unitary matrix whose columns are  bases of the estimated signal subspace. 
We can obtain a better estimate of the channel vector $\bfh\up{\nu}=\bfh[\nu \tau]$ than the one-shot estimate on the current slot, 
by solving the following least-square problem 
\begin{align}
\widehat{\bfw}_\nu=\argmin_{\bfw \in \bC^p} \|\bfx_\nu - \bfB \bfU \bfw\|^2,
\end{align}
from which we can estimate the channel vector by $\widehat{\bfh}_\nu=\bfU \widehat{\bfw}_\nu$. 
If the power of the channel vector $\bfh_\nu$ is not uniformly distributed in different directions spanned by the columns of $\bfU$, this estimate 
can be further improved by weighted least-squares.

In this case, when the channel varies so fast that even the aging over a single slot yields too much degradation of the beamforming performance, 
the multiuser interference can still be managed by exploiting only the subspace information rather than the instantaneous estimate $\widehat{\bfh}_\nu$. For example, the interference from a user with channel vector $\hv_\nu$ can be eliminated by projecting onto the orthogonal complement of its $p$-dim 
subspace. The drawback is that, compared with the projection on the orthogonal complement of $\widehat{\bfh}_\nu$, which wastes only $1$ degree of freedom, 
one wastes $p$ degrees of freedom for zero-forcing a specific user. 
However, this results in a negligible loss when $p \ll M$, especially when a whole group of users spanning roughly the same subspace
can be zero-forced simultaneously \cite{adhikary2013joint,adhikary2014joint}.

It is seen that, in both extreme cases of channel time dynamics, the window of past observations provides very useful information that can be exploited 
at the base-station receiver (uplink) or transmitter (downlink). 
In Section \ref{sec:alg}, we propose an algorithm that uses the training samples $\bfh\up{i}$, $i \in [\nu]$, to find an estimate of 
the $p$-dim signal subspace $\bfU$, which would be exploited in the $\nu$-th training period. 
When this information is used to enhance the channel estimation on the current slot, we evaluate the performance of our algorithm
by looking at the correlation coefficient between the true and the estimated channel vector defined by
\begin{align}\label{eq:eta}
\eta(\bfh_\nu, \widehat{\bfh}_\nu)=\frac{|\inp{\bfh_\nu}{\widehat{\bfh}_\nu}|}{\|\bfh_\nu\| \|\widehat{\bfh}_\nu\|}.
\end{align}
When the subspace information is used to reject interference, we shall look at the normalized residual signal power
\begin{align}\label{eq:mu}
\mu(\hv_\nu, \Um) = \frac{\hv_\nu^\herm (\Id_M - \Um\Um^\herm) \hv_\nu}{\|\hv_\nu\|^2},
\end{align} 
where $\mu(\hv_\nu, \Um)$ measures how much the signal received from a user with channel vector $\bfh_\nu$ can be zero-forced at the uplink receiver when only an estimate of its signal subspace (given by $\bfU$) rather than its channel vector $\bfh_\nu$ is available at the base station. 


\section{Algorithm for Subspace Estimation}\label{sec:alg}
As a robust algorithm for subspace estimation, we use a variant of RMMV (reduced multiple-measurement vector) algorithm that we proposed in \cite{haghighatshoar2015channel}. The main motivation for this algorithm comes from the \textit{multiple measurement vectors (MMV)} problem in compressed sensing. We will briefly explain the MMV problem and why it gives a suitable formulation for subspace estimation in our case. We will also briefly explain the motivation for using RMMV algorithm for extracting the signal subspace.

Consider the channel vectors $\bfh_i$, $i \in [\nu]$, belonging to an observation window of size $T=\nu \tau$. As we explained in Section \ref{sec:ch_model}, we assume that the scattering geometry of the user remains invariant inside this window. This implies that, no matter how the channel dynamics (slowly or quickly varying), the channel vectors of the user inside the window  have a sparse representation in the continuous dictionary $\clA$ consisting of array responses for different AoA $\theta \in [-\theta_{\max}, \theta_{\max}]$. In particular, all the channel vectors $\bfh_i$, $i \in [\nu]$, have the same support in $\clA$, which is given by the AoA $\{\theta_\ell\}_{\ell=1}^p$. This implies that not only every individual channel vector is sparse over $\clA$, but also all the channel vectors together have a joint (group) sparsity structure. This problem has been vastly studied in the compressed sensing literature and it has been shown that exploiting the joint sparsity can further boost the performance, e.g., reduce the number of required measurements (see \cite{tropp2006algorithms, tropp2006algorithms2, lee2012subspace} and the references therein).

Different algorithms have been proposed in the literature for exploiting the joint sparsity such as greedy algorithms \cite{tropp2006algorithms}, convex optimization with a regularization to promote the joint sparsity \cite{tropp2006algorithms2}, subspace methods \cite{lee2012subspace}, and more recent off-grid variants \cite{li2014off, yang2014exact}. In this paper, similar to the one-shot estimation problem \eqref{sparse_alg}, we will focus on atomic norm denoising  for estimating the jointly sparse channel vectors $\bfh_i$, $i\in[\nu]$, from the collection of noisy sketches $\bfx_i=\bfB \bfy_i=\bfB(\bfh_i+\bfn_i)$, $i \in [\nu]$, where the joint sparsity in the channel vectors is incorporated by considering the new dictionary 
\begin{align}
\clD=\{\bfa(\theta) \bfb^\herm: \theta \in [-\theta_{\max}, \theta_{\max}], \bfb \in \bC^\nu\}.
\end{align} 
This approach has been used in \cite{li2014off, yang2014exact}, where it has been shown that, similar to the one-shot variant \eqref{sparse_alg}, the atomic norm denoising can be formulated as an SDP. However, the constraints of this SDP have dimension $(M+\nu)\times (M+\nu)$, which increases by increasing the number of samples. As a result, the computational complexity is quite high even for moderate values $M\approx 64$ and number of samples $\nu \approx 100$.

In \cite{haghighatshoar2015channel}, we proposed the RMMV algorithm, which has nearly the same performance as the SDP proposed in \cite{li2014off, yang2014exact} but its computational complexity does not increase with the sample size $\nu$. This algorithm first computes the sample covariance matrix of $\nu$ samples given by
$\widehat{\bfC}_x=\frac{1}{\nu} \sum_{i\in [\nu]}\bfx_i \bfx_i^\herm$,
 its \textit{singular value decomposition} (SVD) given by $\widehat{\bfC}_x=\bfU \Lambdam \bfU^\herm$, and the low-dimensional data 
given by $\widetilde{\bfX}=\bfU \Lambdam$. It is not difficult to check that $\widetilde{\bfX}=\bfX \bfV_m$, where $\bfX=[\bfx_0, \bfx_1, \dots, \bfx_{\nu-1}]$ is the matrix of the whole sketches, with the SVD $\bfX=\bfU \bfD \bfV^\herm$, where the nonzero singular values in $\bfD$ are the same as nonzero singular values in $\Lambdam$, and where $\bfV_m$ is the $\nu \times m$ matrix consisting of the first $m$ columns of $\bfV$. Note that $\widetilde{\bfX}$ is an $m\times m$ matrix, whose dimension depends on the dimension of the sketches rather than the number of observations $\nu$. It is not difficult to see that, similar to the columns of $\bfX$, the columns of $\widetilde{\bfX}$ still keep their MMV format, i.e., they have the same support over the projected dictionary given by $\bfB \clD=\{\bfB\,\bfa(\theta) \bfb^\herm: \theta \in [-\theta_{\max}, \theta_{\max}], \bfb \in \bC^\nu\}$. The RMMV algorithm is obtained by applying the  atomic norm denoising to the low-dimensional data $\widetilde{\bfX}$, and can be formulated as the following SDP \cite{haghighatshoar2015channel}:
\begin{align}\label{aml_type}
\bfC^*_y =& \argmin_{\Tm \in \bT_+, \bfW\in \bC^{m\times m}} \tr(\bfB \Tm \bfB^\herm) + \tr(\bfW)\nonumber\\
& \text{ subject to } \left [  \begin{array}{cc} \bfB\Tm \bfB^\herm & \widetilde{\bfX} \\ \widetilde{\bfX}^\herm & \bfW \end{array} \right ] \succeq {\bf 0},
\end{align}
where $\bT_+$ denotes the space of all $M\times M$ Hermitian Toeplitz matrices, and where $\bfC^*_y$ is an estimate of the underlying covariance matrix of the whole data samples $\bfy_i=\bfh_i + \bfn_i$, $i\in [\nu]$. Since the array noise is white, the dominant subspace of $\bfC^*_y$ gives an estimate of the signal subspace of $\bfC_h$ (the covariance matrix of the channel vectors). 

\vspace{-1.5mm}
\section{Simulations}
In this section, we assess the performance of our proposed algorithm via numerical simulations. We use $\tau$ as in Section \ref{sec:ch_model} (see Fig.~\ref{fig:pilot}) for the period of training symbols, and $\tau_c$ for the coherence time of the channel. We do simulation for different values of $\tau_c$. When $\tau\approx \tau_c$, the resulting channel vectors are approximately independent from each other, whereas when $\tau \ll \tau_c$, the channel vectors are fully correlated.

\noindent {\bf Channel Model.} We consider a simple model for the channel consisting of $p=3$ multipath components that have equal power with their corresponding AoAs being $\{0,+20,-20\}$ degrees.

\noindent{\bf Array Model and Sampling Scheme.} For simulation, we use an array with $M=64$ antennas. We take $m=16$ orthogonal sketches of the array input signal, thus, the sampling ratio is $\rho=\frac{m}{M}=0.25$. We use an $m\times M$ random binary sampling matrix $\bfB$, which selects $m$ array elements randomly (random antenna selection). In particular, each row of $\bfB$ has only one $1$ is a random antenna location, and has $0$ elsewhere.

\noindent{\bf Window Size.} We use a window of size $\nu=50$, where the signal subspace or the channel vector $\bfh_\nu$ at the last instant $\nu$ is estimated from all the channel vectors $\bfh_i$, $i\in [\nu]$. 

\noindent{\bf Performance Metric.} We consider two performance metrics as explained in Section \ref{exp_past_meas}. When the goal is to use the past observations to enhance the channel estimation on the current slot, we use the correlation coefficient between the true and the estimated channel vector $\eta(\bfh_\nu, \widehat{\bfh}_\nu)$ as defined in \eqref{eq:eta}, and plot the CCDF (complementary cumulative distribution function) of the random variable $20\log_{10}[1/\eta(\bfh_\nu, \widehat{\bfh}_\nu)]$, which is always lower bounded by $0$. Fig.~\ref{fig:eta_measure} shows the simulation results for this case. It is seen that in different regimes of channel variation, i.e., from $\tau\approx \tau_c$ up to $\tau\ll \tau_c$, exploiting the past observations improves the estimation of the channel vector considerably.

\begin{figure*}[h]
\centering
\includegraphics{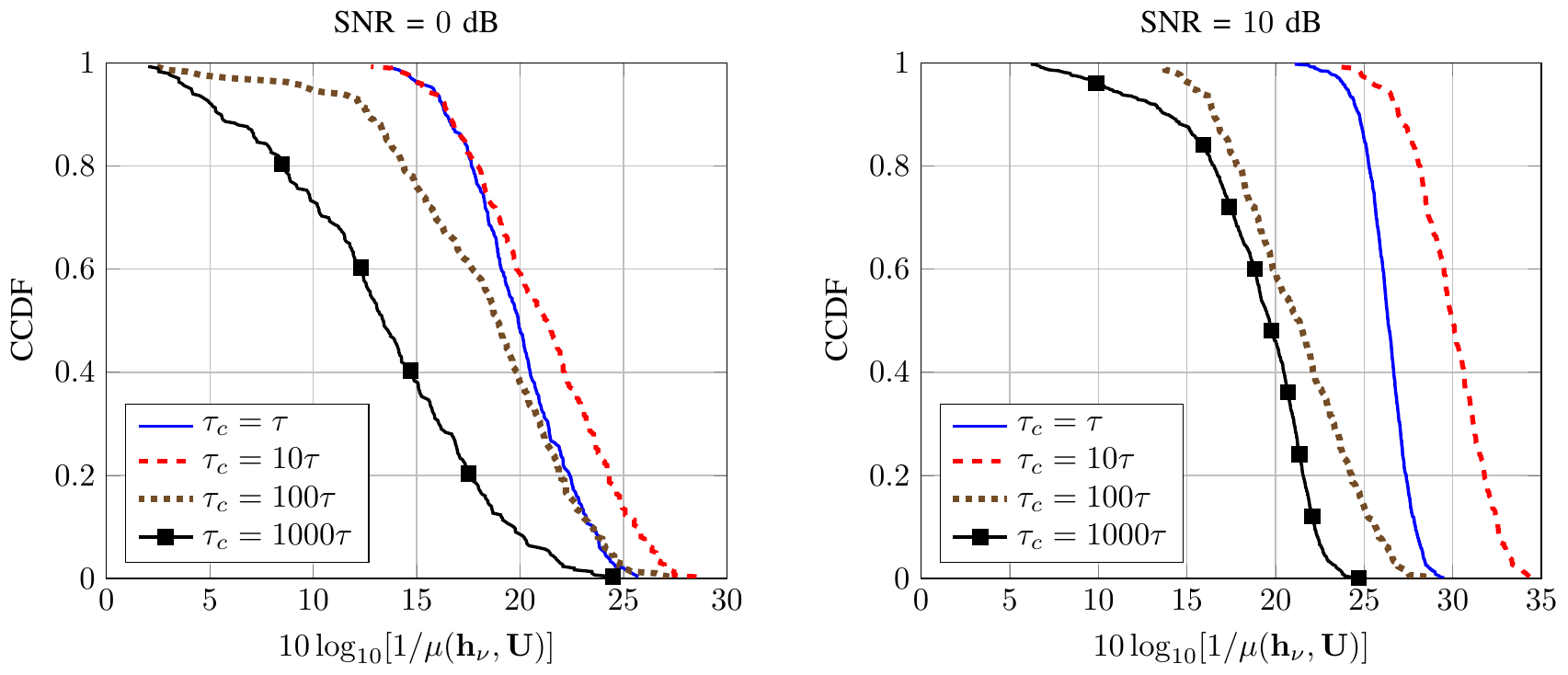}
\caption{The fraction of the power of the $\bfh_\nu$ rejected by projecting onto the estimated subspace for different values of SNR and for different values of channel coherence time $\tau_c$.}
\label{fig:mu_measure}
\end{figure*}

When we use the subspace information to reject interference, we consider normalized residual signal power $\mu(\hv_\nu, \Um)$ defined  by \eqref{eq:mu}, where $\bfU$ is the estimated subspace for the channel vector. We plot the CCDF of the random variable $10\log_{10}[1/\mu(\hv_\nu, \Um)]$ as a performance measure. Fig.~\ref{fig:mu_measure} shows the simulation results. It is again seen that past observations even in a short window of size $\nu=50$, provide a considerable gain in interference rejection for a wide range of SNR and channel variation.

\section{Conclusion}
In this paper, we studied the effect of time-variations of the small-scale fading coefficients due to the large carrier frequency in mm-Wave channels. Inspired by recent results on gridless multiple measurement vectors problem in 
compressed sensing, we proposed an algorithm that exploits the inherent angular sparsity in mm-Wave channels, and the past training symbols in a window of length $T$ to estimate a low-dimensional subspace that approximately contains the channel vector at the current time slot. In particular, our algorithm needs only low-dimensional sketches of the input array signal, and is suitable for HDA implementations. 

We explained that the resulting subspace estimate can be used directly, to separate users in the spatial domain, or indirectly, to improve the estimate of the user channel vector in the current time slot. 
Numerical simulations show very encouraging preliminary results. In particular, they confirm that, due to sparse scattering in mm-Wave channels, the 
channel vector subspace can be robustly estimated for a wide range of channel time-variations and input signal-to-noise ratios even via a short window of past training symbols. Moreover, the extracted subspace information provides a significant improvement  
with respect to the conventional ``one-shot'' techniques. 

\afterpage{\clearpage}
\input{sim_h_estim.txt}

\balance
\newpage
\bibliographystyle{IEEEtran}
\bibliography{references}


\end{document}

%% file: symbols.txt
\def\mindex#1{\index{#1}}

\def\sq{\hbox{\rlap{$\sqcap$}$\sqcup$}}
\def\qed{\ifmmode\sq\else{\unskip\nobreak\hfil
\penalty50\hskip1em\null\nobreak\hfil\sq
\parfillskip=0pt\finalhyphendemerits=0\endgraf}\fi\medskip}

\long\def\defbox#1{\framebox[.9\hsize][c]{\parbox{.85\hsize}{%
\parindent=0pt
\baselineskip=12pt plus .1pt      
\parskip=6pt plus 1.5pt minus 1pt 
 #1}}}

\long\def\beginbox#1\endbox{\subsection*{}%
\hbox{\hspace{.05\hsize}\defbox{\medskip#1\bigskip}}%
\subsection*{}}

\def\endbox{}

\def\tr{\mathsf{Tr}}

\newsavebox{\junk}
\savebox{\junk}[1.6mm]{\hbox{$|\!|\!|$}}

\def\argmin{\mathop{\rm arg\, min}}

\newcommand{\field}[1]{\mathbb{#1}}

\def\ZZ{\field{Z}}

\def\bC{{\mathbb C}}

\def\bE{{\mathbb E}}

\def\bR{{\mathbb R}}

\def\bT{{\mathbb T}}

\def\bfB{{\bf B}}
\def\bfC{{\bf C}}
\def\bfD{{\bf D}}

\def\bfI{{\bf I}}

\def\bfU{{\bf U}}
\def\bfV{{\bf V}}
\def\bfW{{\bf W}}
\def\bfX{{\bf X}}

\def\bfa{{\bf a}}
\def\bfb{{\bf b}}

\def\bfh{{\bf h}}

\def\bfn{{\bf n}}

\def\bfv{{\bf v}}
\def\bfw{{\bf w}}
\def\bfx{{\bf x}}
\def\bfy{{\bf y}}

\def\sfH{{\sf H}}

\def\bfmath#1{{\mathchoice{\mbox{\boldmath$#1$}}%
{\mbox{\boldmath$#1$}}%
{\mbox{\boldmath$\scriptstyle#1$}}%
{\mbox{\boldmath$\scriptscriptstyle#1$}}}}

\def\bfmY{\bfmath{Y}}

\def\bfmhhaY{\bfmath{\hhaY}} 
\def\bfmhhaY{\hbox to 0pt{$\widehat{\bfmY}$\hss}\widehat{\phantom{\raise 1.25pt\hbox{$\bfmY$}}}}

\def\til={{\widetilde =}}

\def\clA{{\cal A}}

\def\clC{{\cal C}}
\def\clD{{\cal D}}

\def\clN{{\cal N}}

\def\clT{{\cal T}}

 \def\FRAC#1#2#3{\genfrac{}{}{}{#1}{#2}{#3}}

\def\ddtp{{\mathchoice{\FRAC{1}{d^{\hbox to 2pt{\rm\tiny +\hss}}}{dt}}%
{\FRAC{1}{d^{\hbox to 2pt{\rm\tiny +\hss}}}{dt}}%
{\FRAC{3}{d^{\hbox to 2pt{\rm\tiny +\hss}}}{dt}}%
{\FRAC{3}{d^{\hbox to 2pt{\rm\tiny +\hss}}}{dt}}}}

\def\average#1,#2,{{1\over #2} \sum_{#1}^{#2}}

\def\eye(#1){{\bf(#1)}\quad}

\def\eq#1/{(\ref{e:#1})}

\newcommand{\inp}[2]{{\langle #1, #2 \rangle}}

\newcommand{\beqn}[1]{\notes{#1}%
\begin{eqnarray} \elabel{#1}}

\newcommand{\eeqn}{\end{eqnarray} }

\newcommand{\beq}[1]{\notes{#1}%
\begin{equation}\elabel{#1}}

\newcommand{\eeq}{\end{equation}}

\def\bdes{\begin{description}}
\def\edes{\end{description}}

\newcounter{rmnum}

\newcounter{anum}

{\end{list}}

\def\ass(#1:#2){(#1\ref{#1:#2})}

\def\ritem#1{
\item[{\sf \ass(\current_model:#1)}]
}

\newenvironment{recall-ass}[1]{%
\begin{description}
\def\current_model{#1}}{
\end{description}
}



%% file: macros.txt
\long\def\comment#1{}

\newfont{\bbb}{msbm10 scaled 700}

\newfont{\bb}{msbm10 scaled 1100}

\newcommand{\av}{{\bf a}}

\newcommand{\hv}{{\bf h}}

\newcommand{\xv}{{\bf x}}

\newcommand{\Bm}{{\bf B}}

\newcommand{\Id}{{\bf I}}

\newcommand{\Tm}{{\bf T}}
\newcommand{\Um}{{\bf U}}

\newcommand{\Xm}{{\bf X}}

\newcommand{\Xc}{{\cal X}}

\newcommand{\Lambdam}{\hbox{\boldmath$\Lambda$}}

\newcommand{\trace}{{\hbox{tr}}}

\newcommand{\transp}{{\sf T}}



%% file: channel_fig.txt
\begin{figure}[h]
\centering
\begin{tikzpicture}[scale=0.65, every node/.style={scale=1}]
\tikzstyle{st_axis}=[thick,dashed,->, blue];
\tikzstyle{st_array}=[line width=2, rotate=45];
\tikzstyle{st_arr_elem}=[line width=1, red];

\draw[red] (0,0) -- (1.5,0);

\draw[thick] (0,-2.4) -- (0,2.4);
\draw[thick, ->] (0,2.4) -- (-0.1,2.4) -- (0.1, 2.6) -- (0,2.6) -- (0,3.5);

\draw (0,-2) node{$\bullet$} node[left] {$0$};
\draw (0,-1) node{$\bullet$} node[left] {$d$};
\draw (0,0) node{$\bullet$} node[left] {$2d$};
\node (BS) at (0,0) {$$};
\draw (0,1) node{$\bullet$} node[left] {$3d$};
\draw (0,2) node{$\bullet$} node[left] {$4d$};
\draw (0,3) node{$\bullet$} node[left] {$(M-1)d$};

\draw[red, thick] (8,-0.25) -- (8,0.25) node[black, above=0.5cm] {User};
\node (U) at (8,0) {$$};
\draw[red, thick] (8,0.25) -- +(45:0.3);
\draw[red, thick] (8,0.25) -- +(135:0.3);

\node[draw, circle, shade] (A) at (4,3) {$$};
\node[draw, circle, shade] (B) at (3.5,1) {$$};
\node[draw, circle, shade] (C) at (4.5,-1) {$$};
\node[draw, circle, shade] (D) at (4,-3) {$$};
\node at (4,0) {$\vdots$};
\draw[very thick, brown, dashed] (4,0) ellipse (1 and 3.5);

\foreach \x in {A,B,C,D}{
	\draw[->] (U) -- (\x) -- (BS);
}

\draw (A) node[above=0.5cm] {Scattering Channel};
\draw[blue, thick,->] (0:1) arc(0:38:1)  node[above] {$\theta_i$};
\end{tikzpicture}
\caption{Scattering channel with discrete angle of arrivals.}
\label{fig:sc_channel}
\end{figure}

%% file: training_fig.txt
\begin{figure}[h]
\centering
\begin{tikzpicture}[scale=1.5, every node/.style={scale=1}]
\draw[thick] (-0.1,0) -- (4.1,0);

\draw[thick,->] (0,0) -- (0,1) node[above] {$s_0$};
\draw[thick,->] (1,0) -- (1,1) node[above] {$s_1$};
\draw[thick,->] (3,0) -- (3,1) node[above] {$s_{\nu-1}$};
\draw[thick,->] (4,0) -- (4,1) node[above] {$s_{\nu}$};

\draw[very thick, dotted] (1.5,0.5) -- (2.5,0.5);
\draw[thick,dashed,<->,red] (0.05,0.5) -- (0.95,0.5) node[midway,above] {$\tau$};
\draw[thick,dashed,<->,red] (3.05,0.5) -- (3.95,0.5) node[midway,above] {$\tau$};

\draw[thick,dashed,<->,blue] (-0.05,-0.2) -- (4.05,-0.2) node[midway,below] {$T$};

\end{tikzpicture}
\caption{Periodic pilot transmission for channel estimation.}
\label{fig:pilot}
\end{figure}

%% file: sim_h_estim.txt
\begin{figure*}[h]
\centering
\begin{tikzpicture}[scale=0.52]
\pgfmathsetmacro{\scale}{1}
\pgfmathsetmacro{\shift}{8.5}
\pgfmathsetmacro{\mksz}{3.5pt}

\pgfplotsset{every axis/.append style={
           legend style={font=\Large,mark size=\mksz},
           label style={font=\large},
           title style={font=\Large},
}}

\begin{scope}[scale=\scale]
\begin{axis}[xlabel={$20\log_{10}[1/\eta(\bfh_\nu, \widehat{\bfh}_\nu)]$},  ylabel={CCDF (SNR = $0$ dB)}, title={$\tau_c=\tau$}, xmin=0, xmax=15, ymax=1, ymin=0, grid=major,  legend style={
        cells={anchor=east},
        legend pos= north east,
    }]
    
\addplot+[no marks, very thick] table[y=prob, x=mmv_loss] {nu50_snr0_tau_settle1_M64_m16.txt};
\addplot+[no marks, line width=2pt, dashed] table[y=prob, x=single_shot_loss] {nu50_snr0_tau_settle1_M64_m16.txt};
\legend{MMV, {One-shot}};
\end{axis}
\end{scope}

\begin{scope}[scale=\scale, xshift=\shift cm]
\begin{axis}[xlabel={$20\log_{10}[1/\eta(\bfh_\nu, \widehat{\bfh}_\nu)]$},  ylabel={}, title={$\tau_c=10 \tau$}, xmin=0, xmax=14, ymax=1, ymin=0, grid=major,  legend style={
        cells={anchor=east},
        legend pos= north east,
    }]
    
\addplot+[no marks, very thick] table[y=prob, x=mmv_loss] {nu50_snr0_tau_settle10_M64_m16.txt};
\addplot+[no marks, line width=2pt, dashed] table[y=prob, x=single_shot_loss] {nu50_snr0_tau_settle10_M64_m16.txt};
\legend{MMV, {One-shot}};
\end{axis}
\end{scope}

\begin{scope}[scale=\scale, xshift=2*\shift cm]
\begin{axis}[xlabel={$20\log_{10}[1/\eta(\bfh_\nu, \widehat{\bfh}_\nu)]$},  ylabel={}, title={$\tau_c=100 \tau$}, xmin=0, xmax=14, ymax=1, ymin=0, grid=major,  legend style={
        cells={anchor=east},
        legend pos= north east,
    }]
    
\addplot+[no marks, very thick] table[y=prob, x=mmv_loss] {nu50_snr0_tau_settle100_M64_m16.txt};
\addplot+[no marks, line width=2pt, dashed] table[y=prob, x=single_shot_loss] {nu50_snr0_tau_settle100_M64_m16.txt};
\legend{MMV, {One-shot}};
\end{axis}
\end{scope}

\begin{scope}[scale=\scale, xshift=3*\shift cm]
\begin{axis}[xlabel={$20\log_{10}[1/\eta(\bfh_\nu, \widehat{\bfh}_\nu)]$},  ylabel={}, title={$\tau_c=1000 \tau$}, xmin=0, xmax=14, ymax=1, ymin=0, grid=major,  legend style={
        cells={anchor=east},
        legend pos= north east,
    }]
    
\addplot+[no marks, very thick] table[y=prob, x=mmv_loss] {nu50_snr0_tau_settle1000_M64_m16_ver2.txt};
\addplot+[no marks, line width=2pt, dashed] table[y=prob, x=single_shot_loss] {nu50_snr0_tau_settle1000_M64_m16_ver2.txt};
\legend{MMV, {One-shot}};
\end{axis}
\end{scope}

\begin{scope}[scale=\scale, yshift=-\shift cm]
\begin{axis}[xlabel={$20\log_{10}[1/\eta(\bfh_\nu, \widehat{\bfh}_\nu)]$},  ylabel={CCDF (SNR = $10$ dB)}, title={}, xmin=0, xmax=4, ymax=1, ymin=0, grid=major,  legend style={
        cells={anchor=east},
        legend pos= north east,
    }]
    
\addplot+[no marks, very thick] table[y=prob, x=mmv_loss] {nu50_snr10_tau_settle1_M64_m16_ver2.txt};
\addplot+[no marks, line width=2pt, dashed] table[y=prob, x=single_shot_loss] {nu50_snr10_tau_settle1_M64_m16_ver2.txt};
\legend{MMV, {One-shot}};
\end{axis}
\end{scope}

\begin{scope}[scale=\scale, xshift=\shift cm, yshift=-\shift cm]
\begin{axis}[xlabel={$20\log_{10}[1/\eta(\bfh_\nu, \widehat{\bfh}_\nu)]$},  ylabel={}, title={}, xmin=0, xmax=4, ymax=1, ymin=0, grid=major,  legend style={
        cells={anchor=east},
        legend pos= north east,
    }]
    
\addplot+[no marks, very thick] table[y=prob, x=mmv_loss] {nu50_snr10_tau_settle10_M64_m16_ver2.txt};
\addplot+[no marks, line width=2pt, dashed] table[y=prob, x=single_shot_loss] {nu50_snr10_tau_settle10_M64_m16_ver2.txt};
\legend{MMV, {One-shot}};
\end{axis}
\end{scope}

\begin{scope}[scale=\scale, xshift=2*\shift cm, yshift=-\shift cm]
\begin{axis}[xlabel={$20\log_{10}[1/\eta(\bfh_\nu, \widehat{\bfh}_\nu)]$},  ylabel={}, title={}, xmin=0, xmax=4, ymax=1, ymin=0, grid=major,  legend style={
        cells={anchor=east},
        legend pos= north east,
    }]
    
\addplot+[no marks, very thick] table[y=prob, x=mmv_loss] {nu50_snr10_tau_settle100_M64_m16.txt};
\addplot+[no marks, line width=2pt, dashed] table[y=prob, x=single_shot_loss] {nu50_snr10_tau_settle100_M64_m16.txt};
\legend{MMV, {One-shot}};
\end{axis}
\end{scope}

\begin{scope}[scale=\scale, xshift=3*\shift cm, yshift=-\shift cm]
\begin{axis}[xlabel={$20\log_{10}[1/\eta(\bfh_\nu, \widehat{\bfh}_\nu)]$},  ylabel={}, title={}, xmin=0, xmax=4, ymax=1, ymin=0, grid=major,  legend style={
        cells={anchor=east},
        legend pos= north east,
    }]
    
\addplot+[no marks, very thick] table[y=prob, x=mmv_loss] {nu50_snr10_tau_settle1000_M64_m16.txt};
\addplot+[no marks, line width=2pt, dashed] table[y=prob, x=single_shot_loss] {nu50_snr10_tau_settle1000_M64_m16.txt};
\legend{MMV, {One-shot}};
\end{axis}
\end{scope}

\end{tikzpicture}
\caption{Comparing the performance of MMV method with the traditional One-shot channel estimation for different SNR and different channel coherence time $\tau_c$. Window size  $\nu=50$, number of array elements $M=64$, dimension of the sketches $m=16$, and sampling scheme is \textit{random antenna selection} (the sampling matrix $\bfB$ is a binary matrix with only one $1$ in each row). }
\label{fig:eta_measure}
\end{figure*}